\documentclass{aastex631}

\usepackage{amsmath}

\begin{document}

\title{Examining the rotation of the planet-hosting M dwarf GJ~3942}




\author[0000-0003-3545-4240]{Andrew Fonseca}
\author[0000-0002-8796-4974]{Sarah Dodson-Robinson}
\affiliation{University of Delaware \\
Sharp Laboratory 217, Newark, DE 19716, USA \\}
\email{afonse@udel.edu, sdr@udel.edu}


\begin{abstract}

Based on radial velocities, EXORAP photometry, and activity indicators, the HADES team reported a 16.3-day rotation period for the M dwarf GJ~3942. However, an RV--H$\alpha$ magnitude-squared coherence estimate has significant peaks at frequencies $1/16$~day$^{-1}$ and $1/32$~day$^{-1}$. We re-analyze HADES data plus {\it Hipparcos}, SuperWASP, and TESS photometry to see whether the rotation period could be 32~days with 16-day harmonic. SuperWASP shows no significant periodicities, while the {\it Hipparcos} observing cadence is suboptimal for detecting 16- and 32-day periodicities. Although the average TESS periodogram has peaks at harmonics of $1/16$~day$^{-1}$, the harmonic sequence is not fully resolved according to the Rayleigh criterion. The TESS observations suggest a $1/16$~day$^{-1}$ rotation frequency and a $1/32$~day$^{-1}$ subharmonic, though resolution makes the TESS rotation detection ambiguous.


\end{abstract}


\section{Introduction} \label{sec:intro}

GJ 3942, an M0.5V star \citep{lepine13} at a distance of 16.9~pc \citep{gaia}, is a HArps-n red Dwarf Exoplanet Survey (HADES) target with 145 high-resolution spectra obtained over a duration of 1203~nights. In addition to the planet GJ~3942~b, indicated by a 6.9-day RV signal with no counterpart in the Ca II S-index and H$\alpha$ activity indicators, \citet{perger17} also found 16.3-day periodicity in the RV, S-index, H$\alpha$, and EXOplanetary systems Robotic APT2 Photometry (EXORAP) time series that they attributed to star rotation.

\section{Magnitude-squared coherence between RVs and activity indicators} \label{sec:RV}

To investigate stellar activity, we estimated magnitude-squared coherence $\hat{C}^2_{xy}(f)$ between RVs and activity indicators. $\hat{C}^2_{xy}(f)$ is a frequency-dependent correlation coefficient between simultaneous time series $x_t$ and $y_t$:
\begin{equation}
    \hat{C}^2_{xy}(f) = \frac{\sum_{k=0}^{K-1} |\hat{S}^{(k)}_{xy}(f)|^2}{\left( \sum_{k=0}^{K-1} \hat{S}^{(k)}_{xx}(f) \right) \left( \sum_{k=0}^{K-1} \hat{S}^{(k)}_{yy}(f) \right)}.
    \label{eq:msc}
\end{equation}
In Equation \eqref{eq:msc}, $f$ is frequency, $x_t$ and $y_t$ are broken into $K$ segments, $\hat{S}^{(k)}_{xx}(f)$ and $\hat{S}^{(k)}_{yy}(f)$ are the estimated power spectra of $x_t$ and $y_t$ from segment $k$, and $\hat{S}^{(k)}_{xy}(f)$ is the estimated cross-spectrum of $x_t$ and $y_t$ from segment $k$ \citep[e.g.][]{pw2020}. A statistically significant peak in $\hat{C}^2_{xy}(f)$ suggests that stellar activity generates the RV oscillation at frequency $f$ \citep{Dodson-Robinson22}.

The top-left panel of Figure \ref{fig:fig1} shows $z(f)$, a \citet{fisher29} transformation of $\hat{C}^2_{xy}(f)$ computed with $x_t = \mathrm{H}\alpha$ index \citep{gomesdasilva11}, $y_t = {\rm RV}$ from the TERRA pipeline \citep{anglada12}, and each time series broken into two 50\% overlapping segments \citep{welch67} with four-term Blackman-Harris tapers applied to each \citep{harris78}. Nearly reaching the 0.1\% false alarm level are shared oscillations at $f = 1/16$~day$^{-1}$, the rotation signal reported by \citet{perger17}, and $f = 1/32$~day$^{-1}$. The $f = 1/32$~day$^{-1}$ signal also appears in $\hat{S}_{yy}(f)$, the RV power spectrum estimate in the denominator of Equation \eqref{eq:msc}.

Rotation often produces significant power at a harmonic of the rotation frequency in addition to the fundamental, especially when the star has multiple spot groups \citep[e.g.][]{boisse11, balona13, ramia18}. Here we explore the possibility that GJ~3942 has a true rotation period of 32~days with a harmonic at 16~days.


\section{Photometry}

In addition to the HARPS-N data, photometric time series from SuperWASP \citep{pollacco06}, {\it Hipparcos} \citep{perryman97, vanleeuwen97}, and TESS \citep{ricker15} are publicly available. (The EXORAP photometry from \citet{perger17} appears not to be publicly available.) We perform Fourier analysis of the SuperWASP, {\it Hipparcos}, and TESS data in order to estimate the star's rotation period.

\subsection{SuperWASP}
\label{subsec:superwasp}

SuperWASP (Wide Angle Search for Planets) is a sky survey based at La Palma and South African Astronomical Observatory. We obtained photometry from NASA Exoplanet Archive, which ingested the WASP data release covering 2004--2008 \citep{butters10}.
The SuperWASP time series contains 7447 observations in a passband with range 400--700 nm \citep{pollacco06}. Consistent with \citet{perger17}, we found no significant signals in the Lomb-Scargle periodogram \citep{lomb76, scargle82}. 

The SuperWASP photometry has a rapid observing cadence with median timestep $4.97 \times 10^{-4}$~s. In an attempt to boost low-amplitude signals with frequencies approximately $1/32$~day$^{-1}$ or $1/16$~day$^{-1}$, we used the \texttt{lightkurve} package to bin the data to a cadence of 1~day. We also tried removing the data taken before and after the most active observing campaign, which covered Julian dates 2454189--2454554, to improve the spectral window. Neither strategy produced a significant peak in the Lomb-Scargle periodogram.

In the top-right panel of Figure \ref{fig:fig1} we visualize a {\it pseudowindow}, which is the frequency-dependent shape of a periodogram peak resulting from a sinusoid placed on the grid of observation times. The pseudowindow shown is an estimated power spectrum of the signal $x_t = \cos (2 \pi f_0 t)$ for $f_0 = 1/16.3$~day$^{-1}$ on the 1-day binned SuperWASP timestamps. The peak at $1/16.3$~day$^{-1}$ demonstrates that the timing of the binned observations is sensitive to 16-day periodicities, but the ground-based SuperWASP photometry does not detect the star rotation.

\subsection{{\it Hipparcos}} \label{sec:hipparcos}

We also analyzed the GJ~3942 observations from the {\it Hipparcos} satellite \citep{perryman97, vanleeuwen97}. The time series consists of 116 data points spanning a time baseline of 1160 days. After clipping a $>3 \sigma$ outlier, we computed a Lomb-Scargle periodogram, which showed no significant oscillations.

To test whether the {\it Hipparcos} observing cadence could detect oscillations with 16- or 32-day periods, we computed pseudowindows for both periods using the procedure described in \S \ref{subsec:superwasp}. The 16-day pseudowindow is pictured in the middle-left panel of Figure \ref{fig:fig1}. Although there is a peak at $f = 1/16$~day$^{-1}$, there is a peak of similar height at $f = 1/36$~day$^{-1}$ (0.0278~day$^{-1}$). A 16-day sinusoid on the {\it Hipparcos} timestamps might therefore appear to have a period of 36~days. In the 32-day pseudowindow, the highest peak is at $f = 1/32$~day$^{-1}$, but there is a significant secondary peak at $f = 1/15$~day$^{-1}$.

The {\it Hipparcos} data do not break the degeneracy between the 16- and 32-day possible rotation periods, as the observation timing is suboptimal for detecting both periods.





\subsection{TESS} \label{sec:TESS}

\begin{figure}
    \centering
    \begin{tabular}{cc}
      \includegraphics[width=0.5\linewidth]{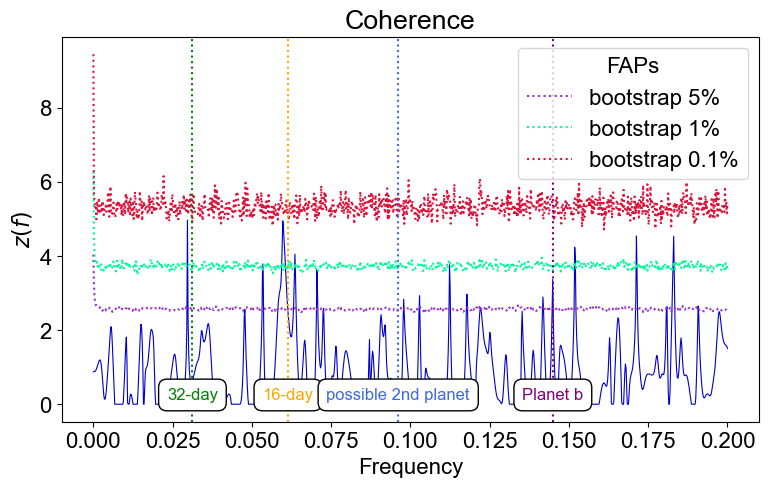} &
      \includegraphics[width=0.5\linewidth]{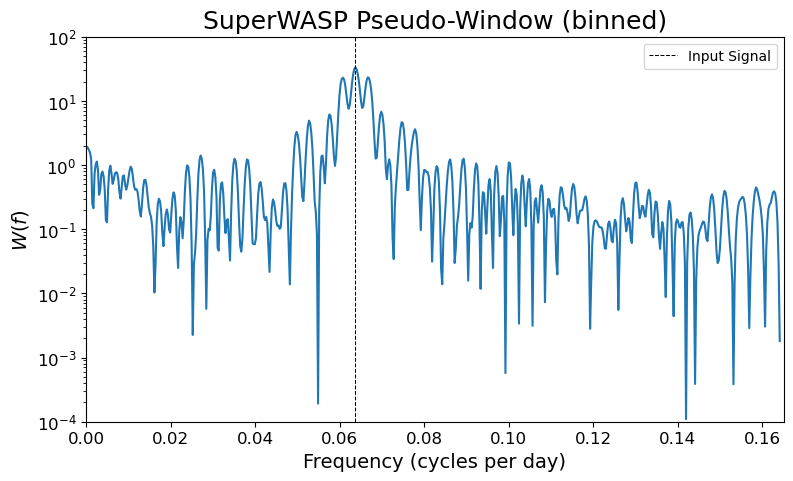} \\
      \includegraphics[width=0.5\linewidth]{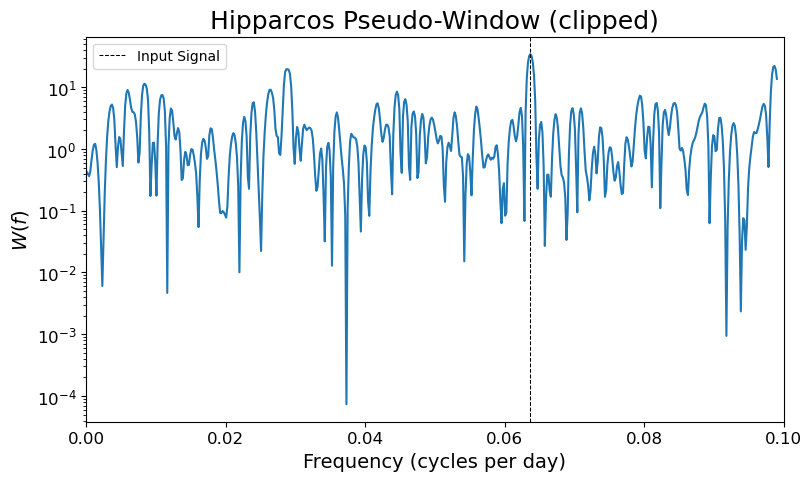} &
      \includegraphics[width=0.5\linewidth]{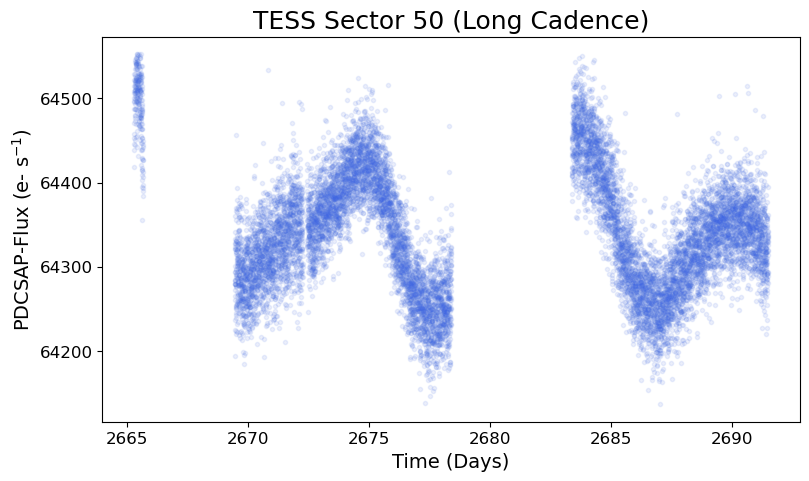} \\
      \includegraphics[width=0.5\linewidth]{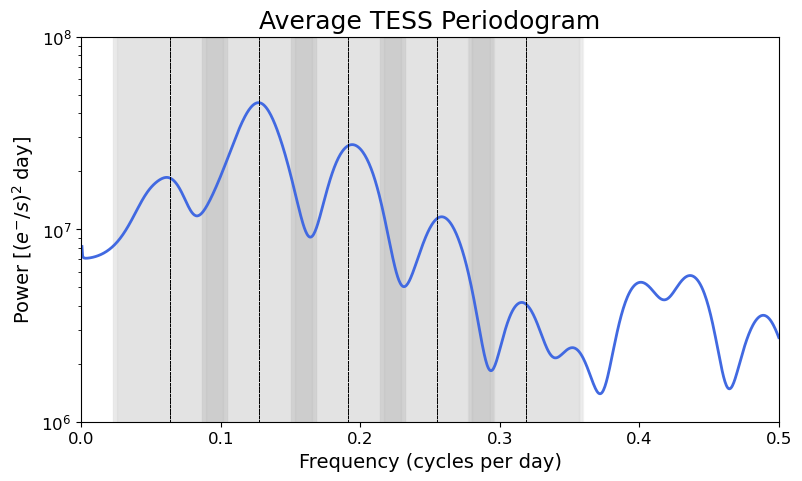} &
      \includegraphics[width=0.5\linewidth]{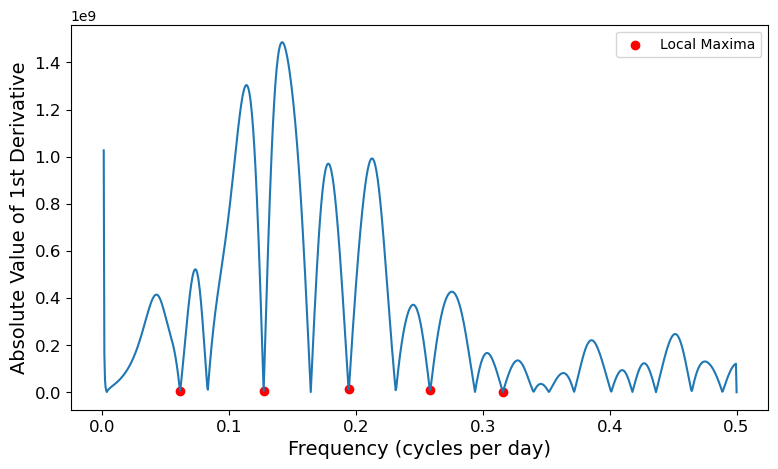}
 \\

    \end{tabular}
    \caption{{\bf Top left:} The Fisher-transformed magnitude-squared coherence estimate from RV and H$\alpha$ has peaks at frequencies $1/32$~day$^{-1}$ and $1/16$~day$^{-1}$. {\bf Top right:} The 16-day SuperWASP pseudowindow after binning to a 1-day cadence. The strong peak shows that the observation timings are sensitive to a 16-day periodicity, though we do not detect it. {\bf Middle left:} The {\it Hipparcos} 16-day pseudowindow lacks a strong peak at $f = 1/16$~day$^{-1}$, indicating that the observation timings are suboptimal for detecting 16-day periodicities. 
    {\bf Middle right:} TESS PDCSAP photometry of GJ~3942 from Sector 52 shows evidence for rotational modulation.
    {\bf Lower left:} Average periodogram from seven TESS light curves. Vertical lines indicate local maxima that may correspond to rotation harmonics. Frequencies within one Rayleigh resolution unit of each local maximum are shaded in light gray.
    {\bf Lower right:} $|d\overline{\mathcal{P}}/df|$ calculated from average TESS periodogram with probable rotation harmonics indicated by red dots.}
    \label{fig:fig1}
\end{figure}

The Transiting Exoplanet Survey Satellite \citep[TESS;][]{ricker15} data are also useful in studies of star rotation and oscillations \citep[e.g.][]{pedersen19, mathys22}. From the MAST archive, we retrieved the 120-second cadence PDCSAP light curves from sectors 23, 50, 51, and 52 \citep{jenkins16} and the 20-second PDCSAP light curves from sectors 50, 51, and 52 \citep{jenkins21}. We clipped $3 \sigma$ flux outliers from each light curve in order to remove periodogram-biasing signals such as flares. The middle-right panel of Figure \ref{fig:fig1} shows the clipped 120-s light curve from Sector 50.

To build a low-variance power spectrum estimate, 
we computed the Lomb-Scargle periodogram of each TESS light curve and then averaged the periodograms, a procedure known as Bartlett's method \citep{barlett48}. The lower-left panel of Figure \ref{fig:fig1} shows the result, in which there is a clear sequence of local maxima. Each maximum is marked by a vertical line, and the frequency range that falls within one Rayleigh resolution unit $\mathcal{R} = 1 / T$ (where $T = 27$~days is the duration of each TESS light curve) of each maximum is shaded in light gray. The dark gray overlap regions cover frequencies within one resolution unit of two different local maxima. The overlaps indicate that the local maxima are not separated in frequency by a full $2 \mathcal{R}$, which means they are not fully resolved from one another \citep{godin72}. However, some numerical experiments suggest that signals are adequately resolved at a frequency separation of $1.45 \mathcal{R}$ \citep[e.g.][]{loumos78, kovacs81}, so we will use the local maxima as the basis for our analysis.

Our hypothesis is that the local maxima in the TESS average periodogram reside at the rotation frequency and its harmonics. To estimate the rotation period, we find the frequencies of the local maxima by using finite differences to approximate $d\overline{\mathcal{P}}/df$, the first derivative of the average periodogram with respect to frequency. At local maxima, $|d\overline{\mathcal{P}}/df| < \epsilon$, where $\epsilon \ll 1$ is a tolerance. The lower-right plot of Figure \ref{fig:fig1} shows $|d\overline{\mathcal{P}}/df|$. Our estimated rotation frequency $f_r$ minimizes
\begin{equation}
    \sum_{i=1}^5 (f_i - i f_r)^2,
    \label{eq:fitrot}
\end{equation}
where $f_i$ are the first five local maxima, marked with red dots in the lower-right panel of Figure \ref{fig:fig1}.

The best-fit rotation period is 15.7~days, which is statistically indistinguishable from the 16.3-day value of \citet{perger17}. Although a 32-day period is undetectable in an average periodogram computed from 27-day light curves, the harmonic sequence in Figure \ref{fig:fig1} nonetheless supports the original rotation period estimate of $\sim 16$~days. Furthermore, when analyzing the time series of concatenated 20-second cadence light curves from Sectors 50--52, in which a 32-day periodicity would be detectable, we do not find a periodogram peak near $1/32$~days.

\section{Conclusion} \label{sec:conclusion}


The GJ~3942 RV time series of \citet{perger17} and the RV--H$\alpha$ magnitude-squared coherence show periodicities at 16~days and 32~days. Here we used SuperWASP, {\it Hipparcos}, and TESS photometry to try and break the degeneracy between the two possible rotation periods. The rotation signal amplitude appears to be too low for a SuperWASP detection, and the {\it Hipparcos} observing cadence is suboptimal for detecting 16- and 32-day periodicities. However, an average periodogram created from seven TESS light curves reveals local maxima at harmonics of 15.7~days, which is statistically indistinguishable from the 16.3-day rotation period reported by \citet{perger17}. The TESS observations suggest that the true rotation period is $\sim 16$~days and the 32-day signal is a subharmonic. However, the TESS rotation detection is ambiguous, as the periodogram peaks that may constitute a harmonic sequence are not fully resolved from one another according to the Rayleigh criterion.


\begin{acknowledgments}

This research used \texttt{Lightkurve}, a \texttt{python} package for Kepler and TESS data analysis (Lightkurve Collaboration, 2018). The time series analyzed in this article are available at \citet{SuperWASP}, \citet{TESS}, \citet{Hipparcos}, \citet{perger17-catalog}. The \texttt{Jupyter} notebooks to execute the analysis in this paper/to generate the figures in this paper is hosted at Github and is preserved on Zenodo at \dataset[doi: 10.5281/zenodo.14187051]{https://doi.org/10.5281/zenodo.14187051} \citep{GJ3942_Time_Series_Analysis_2024}.

This research used the first release of WASP data \citep{butters10} as provided by the WASP consortium and services at the NASA Exoplanet Archive, which is operated by the California Institute of Technology, under contract with the National Aeronautics and Space Administration under the Exoplanet Exploration Program (DOI 10.26133/NEA9).

Work by AF was supported by the NSF's Philadelphia AMP Post-Baccalaureate Research Experience for LSAMP Students Program (grant no.\ 2008197). Work by SDR was supported by NSF grant 2307978.

\end{acknowledgments}

%

\vspace{5mm}
\facilities{HARPS-N, TESS, SuperWASP, Hipparcos}


\software{astropy \citep{astropy13,astropy18}, 
lightkurve
\citep{lightkurve18-software}
NWelch
\citep{Dodson-Robinson22}
}



\bibliography{sample631}{}
\bibliographystyle{aasjournal}



\end{document}